\documentclass[conference]{IEEEtran}
\IEEEoverridecommandlockouts

\usepackage{cite}
\usepackage{amsmath,amssymb,amsfonts}
\usepackage{algorithmic}
\usepackage{graphicx}
\usepackage{textcomp}
\usepackage{xcolor}
\usepackage{url}
\usepackage{booktabs}
\usepackage{multirow}
\usepackage{array}
\usepackage{subcaption}
\usepackage{pifont}

\makeatletter
\def\@seccntformat#1{\csname the#1dis\endcsname\hskip 0.5em}

\makeatother

\makeatletter
\renewcommand{\subsection}{\@startsection{subsection}{2}{\z@}%
  {-3.25ex plus -1ex minus -.2ex}{1.5ex plus .2ex}%
  {\normalfont\bfseries\normalsize}}
\renewcommand{\subsubsection}{\@startsection{subsubsection}{3}{\z@}%
  {-3.25ex plus -1ex minus -.2ex}{1.5ex plus .2ex}%
  {\normalfont\itshape\normalsize}}
\makeatother

\begin{document}

\title{Benchmarking Automatic Speech Recognition for Indian Languages in Agricultural Contexts}

\author{
  \IEEEauthorblockN{Chandrashekar M S, Vineet Singh, and Lakshmi Pedapudi}
  \IEEEauthorblockA{Digital Green\\
  \{chandrashekar, vineet, lakshmi\}@digitalgreen.org}
}

\maketitle

\begin{abstract}
The digitization of agricultural advisory services in India requires robust Automatic Speech Recognition (ASR) systems capable of accurately transcribing domain-specific terminology in multiple Indian languages. This paper presents a benchmarking framework for evaluating ASR performance in agricultural contexts across Hindi, Telugu, and Odia languages. We introduce evaluation metrics including Agriculture Weighted Word Error Rate (AWWER) and domain-specific utility scoring to complement traditional metrics. Our evaluation of 10,934 audio recordings, each transcribed by up to 10 ASR models, reveals performance variations across languages and models, with Hindi achieving the best overall performance (WER: 16.2\%) while Odia presents the greatest challenges (best WER: 35.1\%, achieved only with speaker diarization). We characterize audio quality challenges inherent to real-world agricultural field recordings and demonstrate that speaker diarization with best-speaker selection can substantially reduce WER for multi-speaker recordings (upto 66\% depending on the proportion of multi-speaker audio). We identify recurring error patterns in agricultural terminology and provide practical recommendations for improving ASR systems in low-resource agricultural domains. The study establishes baseline benchmarks for future agricultural ASR development.
\end{abstract}


\section{Introduction}

India's agricultural sector, supporting over 260 million farmers and agricultural workers, faces challenges in accessing timely and accurate agricultural information \cite{agri_stats_2022}. The spread of mobile technology and digital agricultural services has created new opportunities for delivering expert agricultural advisory through voice-based interfaces \cite{aker2016promise}.

Recent advances in deep learning have improved ASR performance for high-resource languages like English \cite{radford2023whisper, babu2021xls}. However, challenges remain for Indian languages, particularly in specialized domains such as agriculture where technical terminology, regional dialects, and acoustic conditions create additional difficulties \cite{bapna2022mslam}.

While general-purpose ASR evaluation frameworks exist, they do not capture the domain-specific requirements of agricultural applications where mistranscribing critical terms like pesticide names or crop diseases can have serious real-world consequences \cite{morris2004mer}. Traditional metrics such as Word Error Rate (WER) treat all words equally, but in agricultural applications errors on domain-critical terminology (e.g.\ pesticide names, crop diseases) should be penalized more heavily than errors on general vocabulary. We therefore introduce AWWER as a task-driven metric design choice (Section~\ref{sec:awwer}) that complements standard error measures such as WER/CER and MER \cite{morris2004mer}.

This paper presents an evaluation framework designed for agricultural ASR systems in Indian languages. Our contributions include:

\begin{itemize}
\item \textbf{Domain-Specific Metrics}: Agriculture Weighted Word Error Rate (AWWER) and LLM-based utility scoring that prioritize domain-critical terms
\item \textbf{Multi-Model Benchmarking}: Evaluation of 10 ASR systems across three Indian languages using multiple evaluation approaches
\item \textbf{Audio Quality Characterization}: Analysis of noise levels and audio issues in real-world agricultural field recordings, revealing the challenges of benchmarking with noisy data
\item \textbf{Speaker Diarization Impact}: Quantification of how speaker diarization and best-speaker selection improve transcription accuracy in multi-speaker agricultural recordings
\item \textbf{Domain Analysis}: Detailed categorization and analysis of agricultural terminology confusion patterns across all three languages
\item \textbf{Practical Guidelines}: Recommendations for model selection and deployment in agricultural contexts
\item \textbf{Open Dataset Release}: Public release of the agricultural ASR benchmark dataset (10,864 audio-transcript pairs across Hindi, Telugu, and Odia) on HuggingFace \cite{digigreen_agri_stt}
\end{itemize}

\section{Related Work}

\subsection{ASR for Indian Languages}

Indian language speech technology has received increasing attention with initiatives like Bhashini \cite{bhashini2022}, Vaani \cite{vaanicitation}, and various academic efforts \cite{dalmia2018sequence, pratap2020mls}. The Vakyansh project \cite{chadha2022vakyansh} has contributed significantly to low-resource Indic language ASR, while Javed et al.\ \cite{javed2022_next_billion} built ASR systems targeting the next billion users with a focus on Indic languages.

Pratap et al. \cite{pratap2020mls} introduced the Multilingual LibriSpeech dataset covering eight languages. More recently, Meta's Massively Multilingual Speech project \cite{meta2023mms} scaled speech recognition to over 1,100 languages, including several Indian languages.

\subsection{Domain-Specific ASR Evaluation}

Domain-specific ASR evaluation has received the most attention in clinical and medical settings, where transcription errors on drug names or dosages carry patient-safety risks. Agricultural applications, by contrast, remain underexplored despite similar stakes: a misrecognized pesticide name or dosage can lead to crop loss or health hazards. To our knowledge, no prior work has proposed domain-weighted error metrics or systematic confusion-pattern analysis for agricultural ASR.

\subsection{ASR Evaluation Metrics}

Traditional ASR metrics include WER, Character Error Rate (CER), and Match Error Rate (MER) \cite{morris2004mer}. However, these metrics have known limitations in domain-specific contexts where all words are treated equally, failing to capture the varying semantic importance of different terms. No existing framework specifically addresses agricultural domain requirements.

\subsection{Agricultural Technology in India}

Digital agricultural platforms in India have grown rapidly, with services like Farmer.Chat (Digital Green's agri-advisory platform), Kisan Call Centre, and various state initiatives \cite{fielke2020digitalisation, rose2021agriculture}. However, language barriers remain a key challenge, and current multilingual ASR systems are not yet accurate enough for reliable use in these settings.

\section{Methodology}

\subsection{Data Collection and Dataset}

Our evaluation dataset comprises 10,934 audio recordings collected from the Farmer.Chat platform between June 2024 and February 2025. The dataset includes:

\begin{itemize}
\item \textbf{Hindi}: 4,626 recordings from Bihar, representing dialectal variations
\item \textbf{Telugu}: 4,075 recordings from Telangana and Andhra Pradesh
\item \textbf{Odia}: 2,233 recordings from Odisha
\end{itemize}

Data was collected from extension workers and farmers using voice queries on mobile devices in real-world agricultural settings. Each recording includes human-annotated reference transcripts and ASR-generated hypothesis transcripts from multiple models.

\subsection{Audio Quality Characteristics}

Agricultural field recordings present audio quality challenges that affect ASR benchmarking. We characterized the noise level and audio issue distribution across all three languages.

\subsubsection{Noise Level Distribution}

Table~\ref{tab:noise_levels} summarizes the noise level distribution. Hindi recordings show the highest proportion of low-noise samples (81.3\%), while Odia has the highest proportion of high-noise recordings (12.3\%), nearly 4.4$\times$ that of Hindi (2.8\%). Telugu falls between the two with 9.2\% high-noise samples.

\begin{table}[t]
\centering
\caption{Noise Level Distribution by Language (\%)}
\label{tab:noise_levels}
\begin{tabular}{@{}lccc@{}}
\toprule
Language & Low & Medium & High \\
\midrule
Hindi  & 81.3 & 15.7 & 3.0 \\
Odia   & 70.5 & 15.9 & 13.6 \\
Telugu & 68.6 & 19.9 & 11.5 \\
\bottomrule
\end{tabular}
\end{table}

\subsubsection{Audio Issue Types}

\begin{figure}[t]
\centering
\includegraphics[width=\columnwidth]{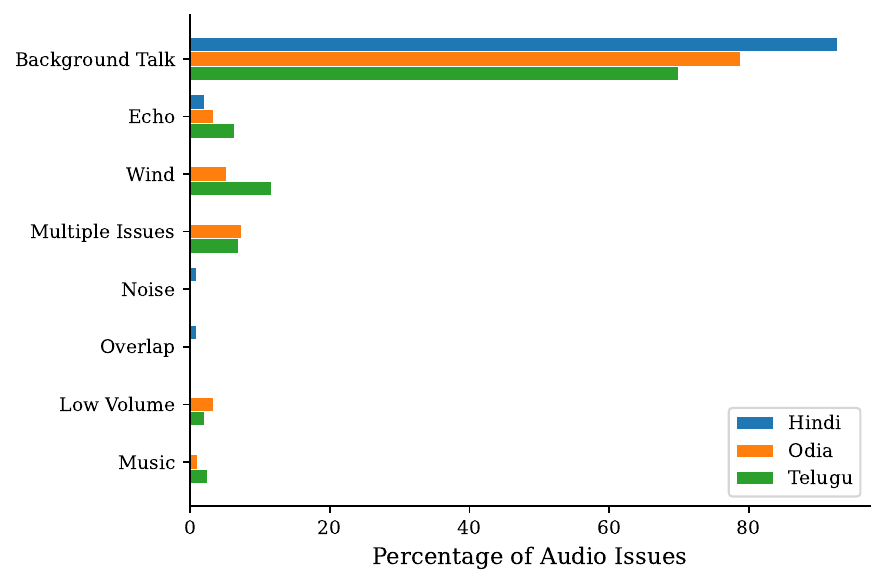}
\caption{Distribution of audio issue types across languages. Background talk dominates all three languages, reflecting real-world agricultural consultation settings.}
\label{fig:audio_issues}
\end{figure}

Fig.~\ref{fig:audio_issues} shows the distribution of audio issues. Background talk is the dominant issue across all languages, reflecting the communal nature of agricultural consultations where multiple people are often present. Hindi exhibits the most granular issue taxonomy compared to Odia and Telugu, partly reflecting the larger Hindi sample size and more detailed annotation. Wind noise is prominent in Telugu and Odia field recordings, while echo and overlap are more common in Hindi, likely due to indoor recording environments.

These quality characteristics reflect the difficulty of obtaining clean benchmarking data from real agricultural field recordings. Traditional training/evaluation speech datasets are often collected in controlled laboratory settings, failing to account for these real issues in production deployments.

\subsection{Evaluation Metrics}

\subsubsection{Traditional Metrics}

\textbf{Word Error Rate (WER)}:
\begin{equation}
WER = \frac{S + D + I}{N}
\end{equation}
where $S$ = substitutions, $D$ = deletions, $I$ = insertions, $N$ = total words in reference

\textbf{Character Error Rate (CER)}:
\begin{equation}
CER = \frac{S_c + D_c + I_c}{C}
\end{equation}
where subscript $c$ denotes character-level operations

\textbf{Match Error Rate (MER)} \cite{morris2004mer}:
\begin{equation}
MER = \frac{S + D + I}{S + D + I + H}
\end{equation}
where $H$ = correct matches

\subsubsection{Agriculture Weighted Word Error Rate (AWWER)}
\label{sec:awwer}

We propose AWWER to account for domain-specific term importance:

\begin{equation}
AWWER = \frac{\sum_{i \in errors} w_i}{\sum_{j \in reference} w_j}
\end{equation}

where $w_i$ represents domain relevance weights assigned via dictionary lookup against a curated agricultural term lexicon:
\begin{itemize}
\item \textbf{Weight 4}: Core agriculture terms (crops, pests, practices)
\item \textbf{Weight 3}: Strongly agriculture-related (soil, weather, timing)
\item \textbf{Weight 2}: Indirectly related (quantities, locations)
\item \textbf{Weight 1}: General vocabulary (default for unmatched tokens)
\end{itemize}

The lexicon is language-specific and derived from human-annotated reference transcripts in the evaluation dataset, supplemented with terms from Farmer.Chat query logs and agricultural extension materials. It is maintained and expanded continuously as new annotated data becomes available. Each reference and hypothesis token is matched against this lexicon to assign its weight. Tokens not found in the lexicon receive weight~1.

\subsubsection{LLM-Based Utility Scoring}
\label{sec:llm_scoring}

We employ OpenAI's GPT-4o to evaluate transcript utility on a 1-4 scale:
\begin{itemize}
\item \textbf{Score 4}: Excellent - minimal difference from reference
\item \textbf{Score 3}: Good - acceptable with same advisory outcome
\item \textbf{Score 2}: Poor - different advisory needed due to key term errors
\item \textbf{Score 1}: Unusable - completely different meaning
\end{itemize}

\subsection{ASR Models Evaluated}

We evaluated 10 ASR systems across different categories. Not all systems support all three languages; Table~\ref{tab:support_matrix} summarizes language coverage. Where a system lacks support for a given language, it is omitted from that language's results table. For systems that support speaker diarization (Gemini 2.5 Pro, Azure Diarize, Google Chirp 3), we report both full-transcript and best-speaker variants.

\textbf{Open Source Models:}
\begin{itemize}
\item Whisper (OpenAI) - API version \cite{radford2023whisper}
\item Meta MMS (Massively Multilingual Speech) \cite{meta2023mms}
\item AI4Bharat IndicASR \cite{javed2022_next_billion}
\end{itemize}

\textbf{Commercial APIs:}
\begin{itemize}
\item Google Speech-to-Text \cite{google2023stt}
\item Google Gemini 2.5 Pro \cite{gemini2025pro} (with speaker diarization)
\item Google Chirp 3 \cite{chirp32025} (with speaker diarization)
\item Microsoft Azure Speech \cite{azure2023speech} (with diarization variant)
\item Sarvam AI \cite{sarvam2023asr}
\end{itemize}

\textbf{Academic/Research Models:}
\begin{itemize}
\item Vaani \cite{vaanicitation} (ASR models from the Vaani speech data initiative, available on HuggingFace)
\item Spring Labs \cite{springinx2023} (SSL and ASR models by SPRING Lab, IIT Madras)
\end{itemize}

\begin{table}[t]
\centering
\caption{Language Support Matrix for Evaluated ASR Models. Models above the mid-rule support all three languages; those below have partial coverage.}
\label{tab:support_matrix}
\footnotesize
\begin{tabular}{@{}lccc@{}}
\toprule
Model & Hindi & Telugu & Odia \\
\midrule
Google STT          & \ding{51} & \ding{51} & \ding{51} \\
Sarvam AI           & \ding{51} & \ding{51} & \ding{51} \\
Azure Speech        & \ding{51} & \ding{51} & \ding{51} \\
MMS (Meta)          & \ding{51} & \ding{51} & \ding{51} \\
Gemini 2.5 Pro      & \ding{51} & \ding{51} & \ding{51} \\
\midrule
Whisper (API)$^a$   & \ding{51} & \ding{55} & \ding{51} \\
Google Chirp 3$^b$  & \ding{51} & \ding{55} & \ding{55} \\
AI4Bharat$^c$       & \ding{51} & \ding{55} & \ding{51} \\
Vaani$^d$           & \ding{51} & \ding{51} & \ding{55} \\
Spring Labs$^d$     & \ding{51} & \ding{51} & \ding{55} \\
\bottomrule
\end{tabular}
\vspace{2pt}
\raggedright\scriptsize

$^a$Telugu output incoherent.

$^b$Telugu/Odia not supported at the time of evaluation.

$^c$Telugu not evaluated.

$^d$Odia not evaluated.
\end{table}

\section{Results and Analysis}

\subsection{Overall Performance by Language}

Table~\ref{tab:overall_performance} presents the best performance achieved across all models for each language. For models that support speaker diarization, we report their best-speaker variant (see Section~\ref{sec:diarization}).

\begin{table}[t]
\centering
\caption{Overall Best Performance by Language}
\label{tab:overall_performance}
\scriptsize
\begin{tabular}{@{}lclcl@{}}
\toprule
Language & Best WER & Model & Best AWWER & Model \\
\midrule
Hindi  & 16.2\% & Google STT & 13.3\% & Gemini 2.5 Pro (BS) \\
Telugu & 33.2\% & Google STT & 28.7\% & Google STT \\
Odia   & 35.1\% & Azure Diar. (BS) & 29.8\% & Azure Diar. (BS) \\
\bottomrule
\end{tabular}
\end{table}

Hindi achieves the lowest WER, consistent with its status as a higher-resource language relative to Telugu and Odia. Odia, while the weakest-performing language in our dataset, shows marked improvement when diarization-enabled models are used, with Azure Diarize (Best Speaker) achieving 35.1\% WER compared to 70.7\% from Google STT without diarization. Telugu maintains intermediate performance with Google Speech-to-Text leading on both WER and AWWER.

\subsection{Model-Specific Performance}

For models that support speaker diarization (Gemini 2.5 Pro, Azure Diarize, Google Chirp 3), we report only the best-speaker variant in the rankings below, as these mostly outperform full-transcript results (see Section~\ref{sec:diarization}). Tables~\ref{tab:hindi_results}--\ref{tab:odia_results} present rankings retaining only the best variant per underlying model.

\subsubsection{Hindi Language Results}

\begin{table}[t]
\centering
\caption{Hindi ASR Model Performance (Ranked by WER)}
\label{tab:hindi_results}
\scriptsize
\begin{tabular}{@{}rlcccc@{}}
\toprule
Rank & Model & WER & CER & MER & AWWER \\
\midrule
1 & Google STT & \textbf{16.2\%} & 12.5\% & 14.4\% & 24.5\% \\
2 & Vaani & 16.6\% & 11.6\% & 14.1\% & 14.4\% \\
3 & Gemini 2.5 Pro (BS) & 18.5\% & 15.8\% & 14.5\% & \textbf{13.3\%} \\
4 & Azure Diarize (BS) & 25.6\% & 19.1\% & 19.4\% & 19.4\% \\
5 & Sarvam AI & 28.7\% & 21.1\% & 21.2\% & 32.5\% \\
6 & Chirp 3 (BS) & 30.0\% & 23.3\% & 22.7\% & 22.7\% \\
7 & Whisper (OpenAI API) & 46.8\% & 32.9\% & 34.0\% & 43.0\% \\
8 & Spring Labs & 47.1\% & 33.2\% & 34.1\% & 38.8\% \\
9 & MMS (Meta) & 51.4\% & 51.0\% & 39.4\% & 49.3\% \\
10 & AI4Bharat & 59.8\% & 52.9\% & 37.9\% & 39.4\% \\
\bottomrule
\end{tabular}
\end{table}

Google Speech-to-Text achieves the lowest WER (16.2\%) for Hindi, closely followed by Vaani (16.6\%) and Gemini 2.5 Pro (Best Speaker) (18.5\%). The top five models achieve WER below 30\%.

\subsubsection{Telugu Language Results}

\begin{table}[t]
\centering
\caption{Telugu ASR Model Performance (Ranked by WER)}
\label{tab:telugu_results}
\scriptsize
\begin{tabular}{@{}rlcccc@{}}
\toprule
Rank & Model & WER & CER & MER & AWWER \\
\midrule
1 & Google STT & \textbf{33.2\%} & 16.2\% & 28.8\% & \textbf{28.7\%} \\
2 & Gemini 2.5 Pro (BS) & 37.9\% & 19.1\% & 30.7\% & 30.2\% \\
3 & Vaani & 38.7\% & 16.4\% & 31.1\% & 30.3\% \\
4 & Azure Diarize (BS) & 41.8\% & 16.3\% & 37.1\% & 37.2\% \\
5 & Spring Labs & 42.2\% & 17.6\% & 34.1\% & 52.0\% \\
6 & Sarvam AI & 46.5\% & 24.2\% & 32.0\% & 32.5\% \\
7 & MMS (Meta) & 67.5\% & 36.7\% & 56.3\% & 71.1\% \\
\bottomrule
\end{tabular}
\end{table}

Telugu performance is led by Google Speech-to-Text at 33.2\% WER. Gemini 2.5 Pro (Best Speaker) is competitive at 37.9\%, followed by Vaani (38.7\%), Azure Diarize (Best Speaker) at 41.8\%, Spring Labs (42.2\%), and Sarvam AI at 46.5\%. Despite its higher WER, Sarvam AI achieves a notably low AWWER of 32.5\%, suggesting relatively good preservation of agricultural terminology (see Section~\ref{sec:awwer_analysis}). The gap between the top model and MMS (67.5\%) is substantial, indicating wide variance in Telugu support quality.
\subsubsection{Odia Language Results}

\begin{table}[t]
\centering
\caption{Odia ASR Model Performance (Ranked by WER)}
\label{tab:odia_results}
\scriptsize
\begin{tabular}{@{}rlcccc@{}}
\toprule
Rank & Model & WER & CER & MER & AWWER \\
\midrule
1 & Azure Diarize (BS) & \textbf{35.1\%} & 15.2\% & 31.6\% & \textbf{29.8\%} \\
2 & Sarvam AI & 35.8\% & 15.3\% & 31.8\% & 43.3\% \\
3 & Gemini 2.5 Pro (BS) & 37.8\% & 22.5\% & 32.9\% & 32.0\% \\
4 & MMS (Meta) & 61.3\% & 35.7\% & 55.4\% & 67.4\% \\
5 & AI4Bharat & 64.8\% & 25.8\% & 50.3\% & 50.8\% \\
6 & Google STT & 70.7\% & 32.6\% & 53.6\% & 54.6\% \\
7 & Whisper (OpenAI API) & 125.6\% & 122.7\% & 100.0\% & 105.3\% \\
\bottomrule
\end{tabular}
\end{table}

Odia results show a different pattern: Azure Diarize (Best Speaker) leads at 35.1\% WER, a large reduction compared to Google STT's 70.7\% WER for Odia (see Table~\ref{tab:odia_results}), demonstrating the impact of diarization for low-resource languages. Sarvam AI and Spring Labs are close behind, both at 35.8\%, followed by Gemini 2.5 Pro (Best Speaker) at 37.8\%. AI4Bharat achieves 64.8\% WER. Google STT, which leads in Hindi and Telugu, drops to 70.7\% WER for Odia, highlighting the low-resource challenge. Whisper (OpenAI API) exceeds 100\% WER (125.6\%), indicating severe difficulty with Odia where insertions outnumber correct words.

\subsection{Impact of Speaker Diarization}
\label{sec:diarization}

Several models in our evaluation support speaker diarization, which identifies and separates speech from different speakers. Multi-speaker recordings are common in agricultural field settings where extension workers consult with groups of farmers. In such cases, diarization enables selecting the transcript of the primary speaker, reducing noise from background conversations.

Table~\ref{tab:diarization} presents the impact of best-speaker selection across models and languages.

\begin{table}[t]
\centering
\caption{Impact of Speaker Diarization: Full Transcript vs Best Speaker WER}
\label{tab:diarization}
\scriptsize
\begin{tabular}{@{}llccccc@{}}
\toprule
Model & Lang. & Full WER & BS WER & Improv. & Avg. Spk. & Multi-Spk.\% \\
\midrule
Google Chirp 3 & Hindi & 88.3\% & 30.0\% & 66.0\% & 1.75 & 56.6\% \\
Gemini 2.5 Pro & Hindi & 53.5\% & \textbf{18.5\%} & \textbf{65.4\%} & 1.37 & 29.2\% \\
Azure Diarize & Hindi & 31.0\% & 25.6\% & 17.5\% & 1.07 & 7.1\% \\
\midrule
Gemini 2.5 Pro & Odia & 47.4\% & 37.8\% & 20.1\% & 1.21 & 17.2\% \\
Azure Diarize & Odia & 35.1\% & \textbf{35.1\%} & $-$0.2\% & 1.02 & 2.2\% \\
\midrule
Gemini 2.5 Pro & Telugu & 57.4\% & \textbf{37.9\%} & \textbf{33.9\%} & 1.17 & 13.8\% \\
Azure Diarize & Telugu & 44.5\% & 41.8\% & 6.1\% & 1.03 & 3.1\% \\
\bottomrule
\end{tabular}
\end{table}

\begin{figure}[t]
\centering
\includegraphics[width=\columnwidth]{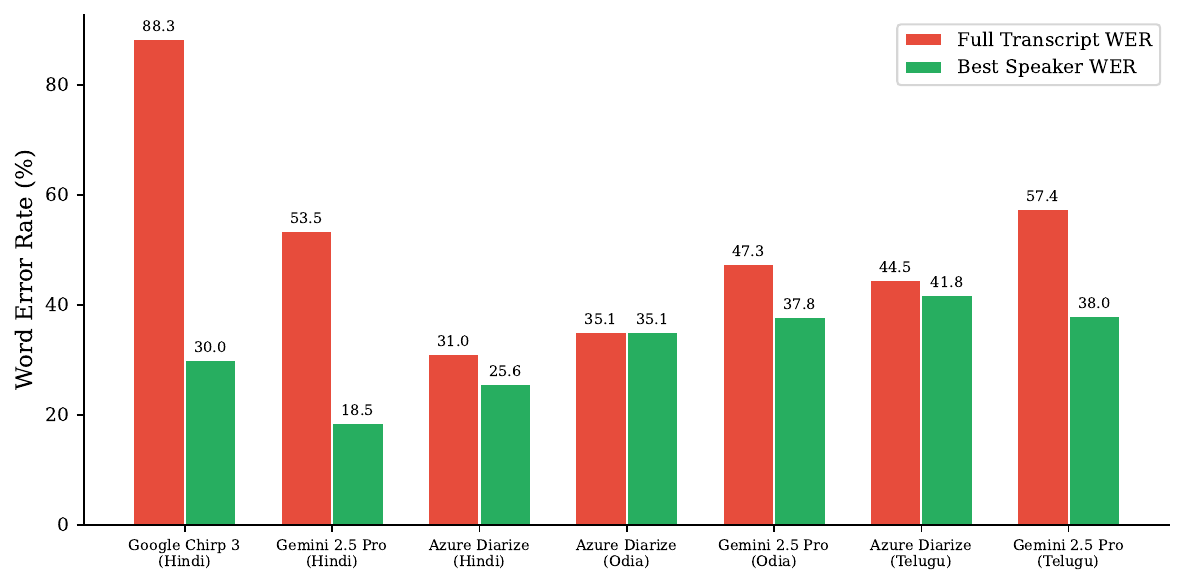}
\caption{Full transcript WER vs best-speaker WER across models and languages. Models with higher multi-speaker percentages show larger improvements from best-speaker selection.}
\label{fig:diarization}
\end{figure}

Key observations:
\begin{itemize}
\item \textbf{Gemini 2.5 Pro} shows the most consistent diarization benefit: 65.4\% improvement in Hindi, 33.9\% in Telugu, and 20.1\% in Odia. This correlates with the multi-speaker percentage in each language's recordings.
\item \textbf{Google Chirp 3} achieves 66.0\% improvement in Hindi (the only language with Chirp 3 data), where 56.6\% of recordings contain multiple speakers.
\item \textbf{Azure Diarize} shows modest improvement except in Odia ($-$0.2\%), where only 2.2\% of recordings contain multiple speakers, making diarization unnecessary.
\item The improvement magnitude correlates strongly with the multi-speaker percentage, confirming that diarization benefit is driven by the presence of extraneous speakers.
\end{itemize}

Based on these findings, all subsequent analyses use the best-speaker variant for models with diarization capability, as it represents the effective performance achievable through readily available post-processing.

\subsection{Agriculture Domain Error Analysis}

\subsubsection{Agricultural Domain Categories}

To analyze agricultural term errors, we defined 12 primary domain categories:

\begin{enumerate}
\item \textbf{Agricultural Practice} - Farming methods and techniques
\item \textbf{Crop Names} - Specific crop varieties
\item \textbf{Pest/Disease} - Plant health issues
\item \textbf{Soil Nutrient/Fertilizer} - Soil management terms
\item \textbf{Chemical Names} - Pesticides and chemicals
\item \textbf{Agriculture Units} - Measurement units
\item \textbf{Season Names} - Agricultural seasons
\item \textbf{Variety Names} - Crop varieties and cultivars
\item \textbf{Symptoms} - Plant disease symptoms
\item \textbf{Month Names} - Seasonal timing references
\item \textbf{Numerals} - Quantities and measurements
\item \textbf{Country/Place} - Geographic references
\end{enumerate}

\subsubsection{Confusion Pattern Analysis}

\begin{figure*}[t]
\centering
\includegraphics[width=\textwidth]{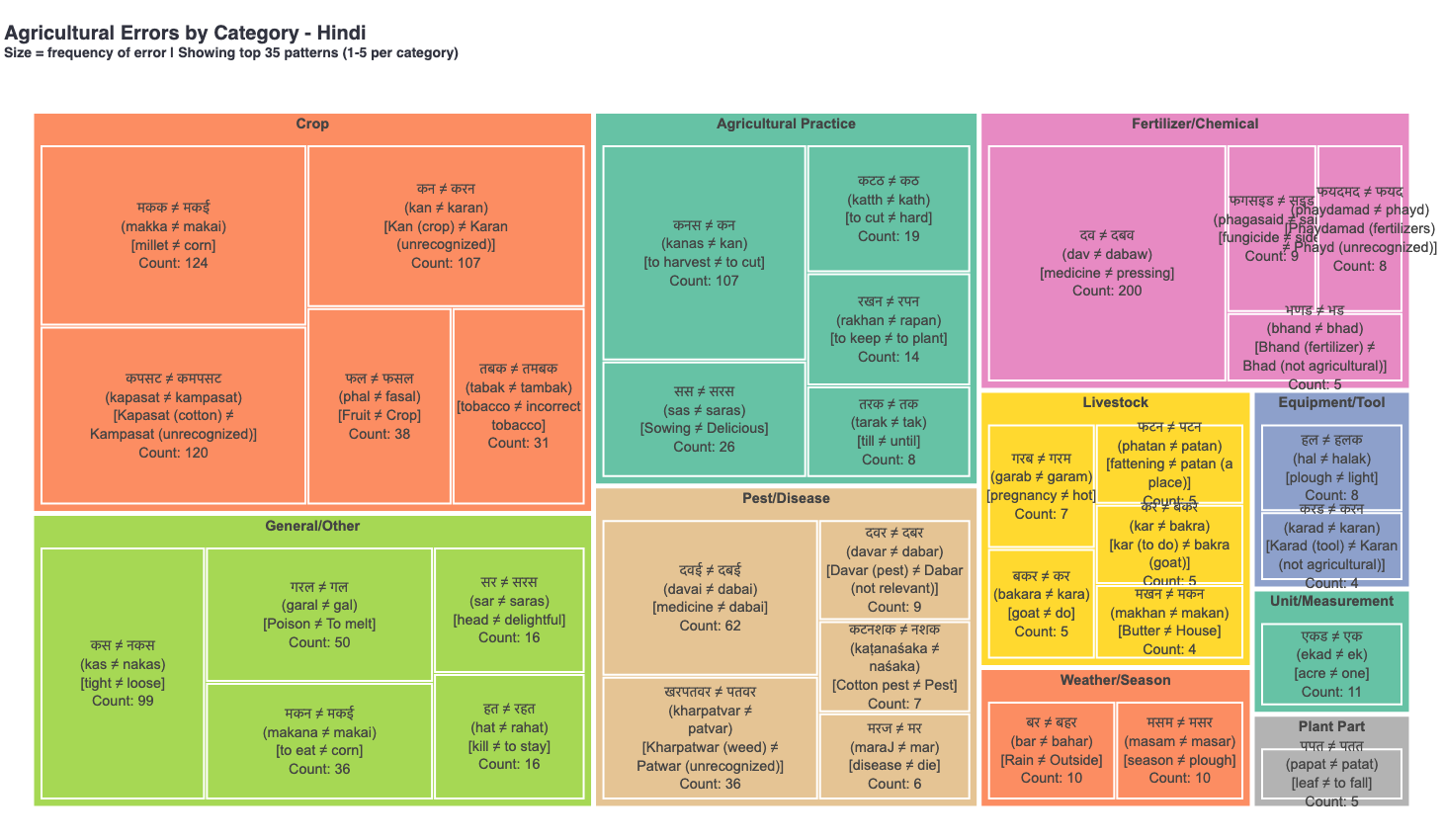}
\caption{Hindi: Treemap of top agricultural term confusion pairs by domain category. Box sizes represent error frequency, with colors indicating domain categories.}
\label{fig:treemap_hindi}
\end{figure*}

\begin{figure*}[t]
\centering
\includegraphics[width=\textwidth]{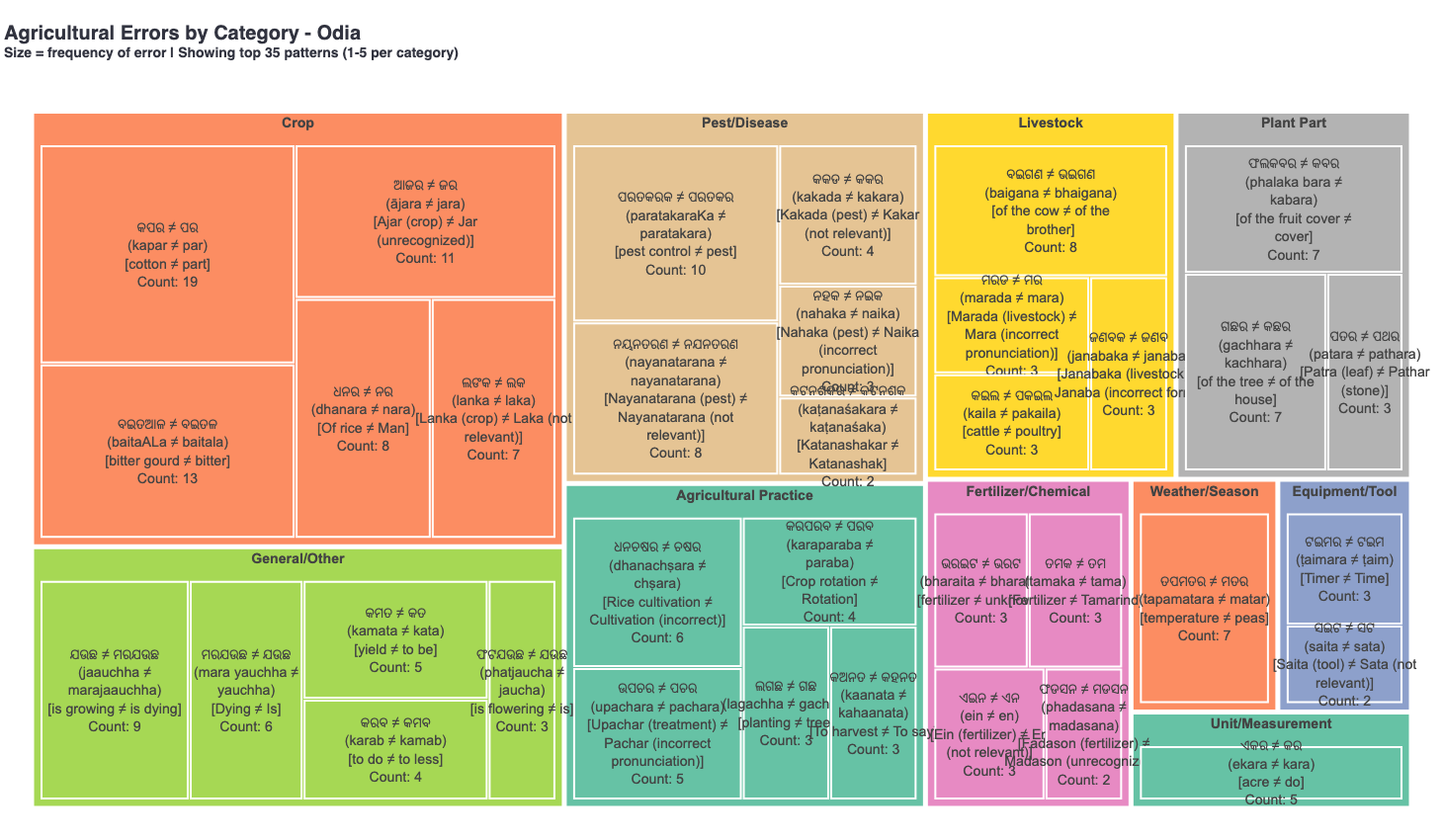}
\caption{Odia: Treemap of top agricultural term confusion pairs by domain category. Box sizes represent error frequency, with colors indicating domain categories.}
\label{fig:treemap_odia}
\end{figure*}

\begin{figure*}[t]
\centering
\includegraphics[width=\textwidth]{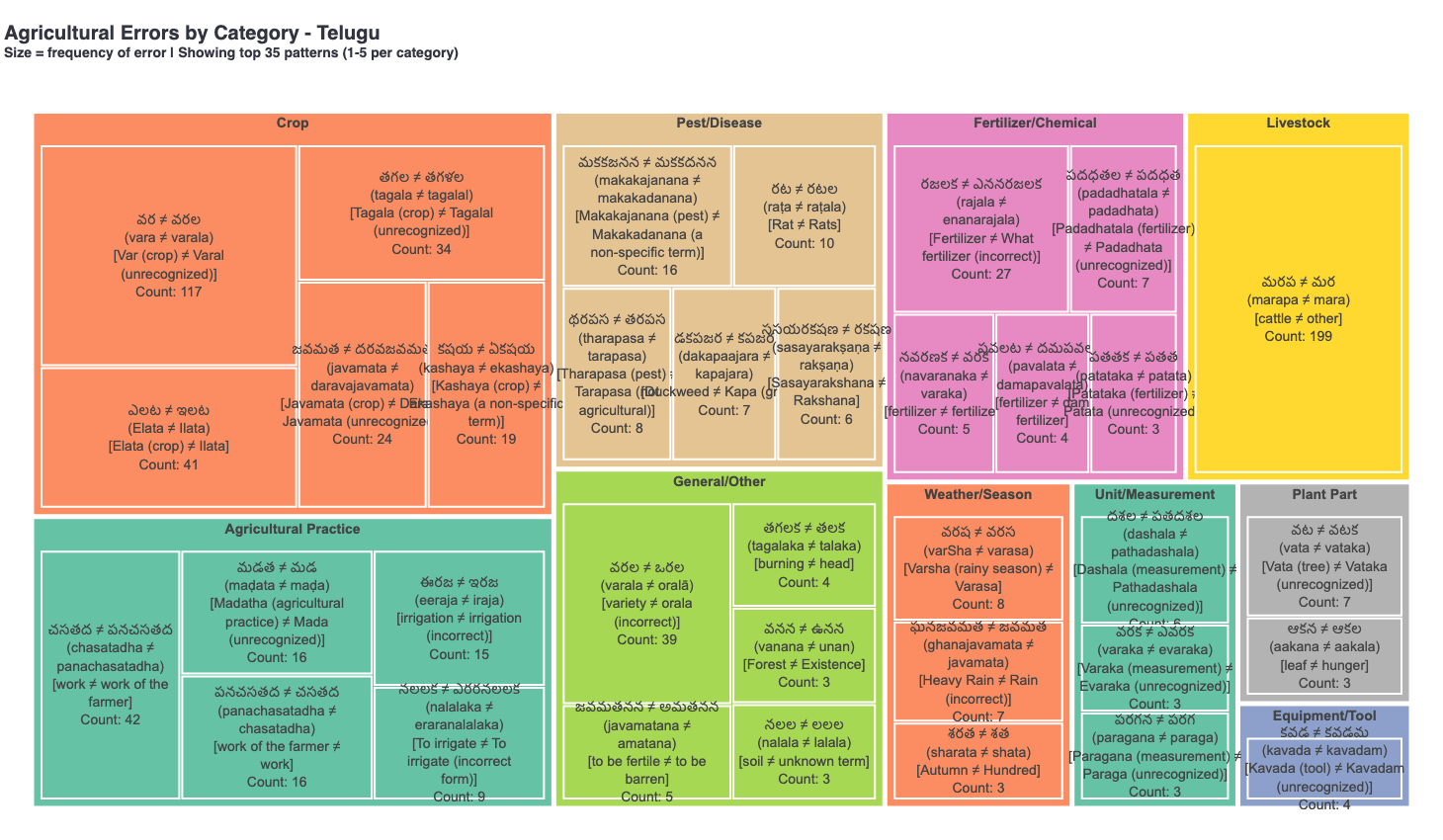}
\caption{Telugu: Treemap of top agricultural term confusion pairs by domain category. Box sizes represent error frequency, with colors indicating domain categories.}
\label{fig:treemap_telugu}
\end{figure*}

Analysis of confusion patterns reveals systematic errors in agricultural terminology. Figs.~\ref{fig:treemap_hindi}--\ref{fig:treemap_telugu} present treemap visualizations of the most frequent confusion pairs across the three languages, categorized by agricultural domain.

Across all three languages, the dominant confusion pairs involve crop names and chemical/fertilizer terms. The specific pairs differ by language and script, but the underlying pattern is consistent: ASR models substitute domain-critical terms with phonetically similar but semantically different words.

\subsubsection{Most Frequent Agriculture Domain Confusions}

Analyzing the most frequent substitution pairs across languages, we observe that high-frequency errors often involve (i) crop names, (ii) chemical/fertilizer terms, (iii) pest/disease vocabulary, and (iv) units/measurement.

\textbf{Hindi (selected high-frequency substitutions):}
\begin{itemize}
\item \textit{dava} $\rightarrow$ \textit{dabav} (Fertilizer/Chemical, 200)
\item \textit{makka} $\rightarrow$ \textit{makai} (Crop, 124)
\item \textit{ekad} $\rightarrow$ \textit{ek} (Unit/Measurement, 11)
\end{itemize}

\textbf{Odia (selected high-frequency substitutions):}
\begin{itemize}
\item \textit{kapura} $\rightarrow$ \textit{pura} (Crop, 19)
\item \textit{baitaalu} $\rightarrow$ \textit{baitalu} (Crop, 13)
\item \textit{pratikarak} $\rightarrow$ \textit{pratikar} (Pest/Disease, 10)
\end{itemize}

\textbf{Telugu (selected high-frequency substitutions):}
\begin{itemize}
\item \textit{tagallu} $\rightarrow$ \textit{tagalla} (Crop, 58)
\item \textit{eeroju} $\rightarrow$ \textit{roju} (Agricultural Practice, 34)
\item \textit{madulu} $\rightarrow$ \textit{madu} (Fertilizer/Chemical, 17)
\end{itemize}

\subsubsection{Error Impact Analysis}

Aggregating errors across all models and languages using AWWER scoring, we found that:
\begin{itemize}
\item 34\% of errors affected high-weight agricultural terms (weight 3--4)
\item 28\% affected medium-weight terms (weight 2)
\item 38\% affected general vocabulary (weight 1)
\end{itemize}

Critical errors like \textit{gehun} $\rightarrow$ \textit{gaon} (wheat $\rightarrow$ village) demonstrate how traditional WER underestimates practical impact in agricultural contexts.

\subsection{AWWER Results Analysis}
\label{sec:awwer_analysis}

AWWER scores reveal domain-specific performance patterns not captured by traditional WER. Table~\ref{tab:awwer_top3} shows the top 3 models per language ranked by AWWER, alongside their WER ranking for comparison.

\begin{table}[t]
\centering
\caption{Top 3 Models by AWWER per Language}
\label{tab:awwer_top3}
\scriptsize
\begin{tabular}{@{}llcccc@{}}
\toprule
Language & Model & WER & AWWER & $\Delta$ & Rank $\Delta$ \\
\midrule
\multirow{3}{*}{Hindi}
 & Gemini 2.5 Pro (BS) & 18.5\% & 13.3\% & $-$5.2 & +2 \\
 & Vaani & 16.6\% & 14.4\% & $-$2.2 & --- \\
 & Azure Diarize (BS) & 25.6\% & 19.4\% & $-$6.2 & +1 \\
\midrule
\multirow{3}{*}{Odia}
 & Azure Diarize (BS) & 35.1\% & 29.8\% & $-$5.3 & --- \\
 & Spring Labs & 35.8\% & 30.3\% & $-$5.5 & +1 \\
 & Gemini 2.5 Pro (BS) & 37.8\% & 32.0\% & $-$5.8 & +1 \\
\midrule
\multirow{3}{*}{Telugu}
 & Google STT & 33.2\% & 28.7\% & $-$4.5 & --- \\
 & Gemini 2.5 Pro (BS) & 37.9\% & 30.2\% & $-$7.7 & --- \\
 & Vaani & 38.7\% & 30.3\% & $-$8.4 & --- \\
\bottomrule
\end{tabular}
\end{table}

\begin{figure}[t]
\centering
\includegraphics[width=\columnwidth]{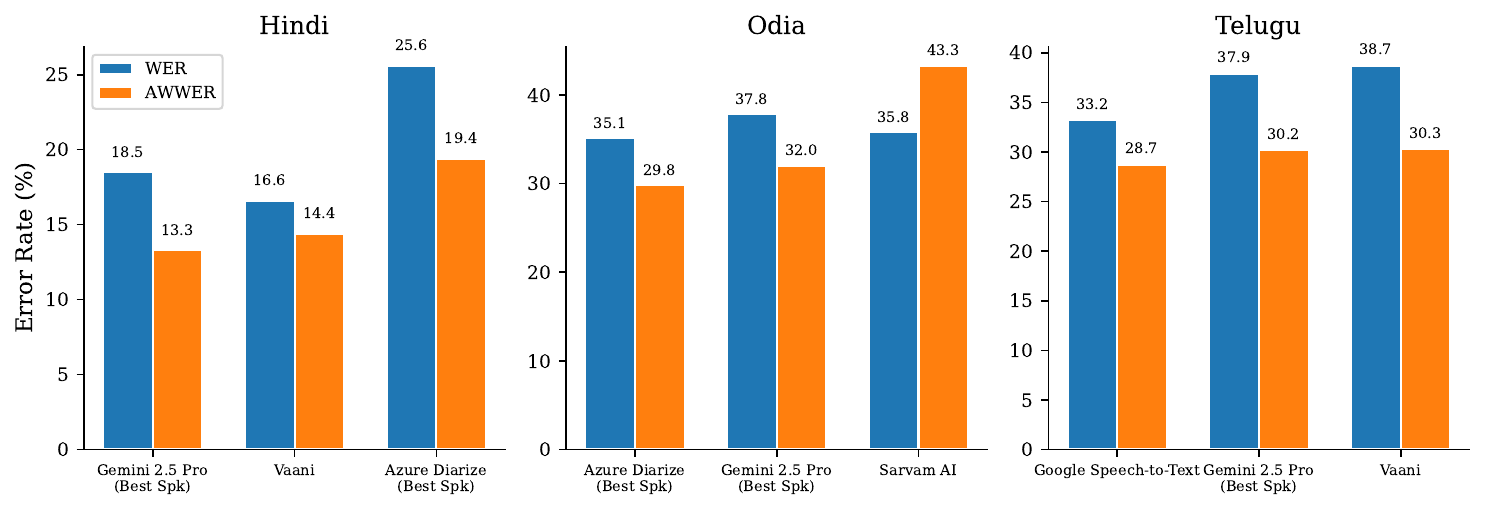}
\caption{WER vs AWWER comparison for top 3 models per language. Models with AWWER $<$ WER (negative delta) concentrate errors on general vocabulary rather than on critical agricultural terms.}
\label{fig:awwer_vs_wer}
\end{figure}

The AWWER-WER delta reveals which models handle agricultural terminology relatively well versus poorly:

\begin{itemize}
\item \textbf{Negative delta} (AWWER $<$ WER): The model makes proportionally fewer errors on domain-critical agricultural terms than on general vocabulary. This is desirable for agricultural applications. Gemini 2.5 Pro (Best Speaker) consistently shows negative delta across all three languages.

\item \textbf{Positive delta} (AWWER $>$ WER): The model makes disproportionately more errors on critical agricultural terms, which is worse for practical utility even if overall WER is acceptable. For example, Sarvam AI in Odia has a competitive WER of 35.8\% but AWWER of 43.3\% ($\Delta$ = +7.5 pp), indicating its errors concentrate on important agricultural vocabulary. Interestingly, Sarvam AI shows the opposite pattern in Telugu: WER of 46.5\% but AWWER of 32.5\% ($\Delta$ = $-$14.0 pp), meaning it handles Telugu agricultural terms relatively well despite higher overall error rates. This cross-language contrast suggests that model performance on domain vocabulary is language-dependent and cannot be assumed from a single language's results.
\end{itemize}

For example, in Hindi: Google Speech-to-Text has the best WER (16.2\%) but its AWWER rank drops to 5th (delta +8.3 pp), meaning it makes relatively more errors on agricultural terms. In contrast, Gemini 2.5 Pro (Best Speaker) ranks 1st in AWWER despite being 3rd in WER, demonstrating superior agricultural term recognition.

\subsection{LLM-Based Utility Evaluation}

We scored the transcripts of the top-performing model for each language (Google STT for Hindi and Telugu, Azure Diarize Best Speaker for Odia) using GPT-4o. For each transcript, the model was prompted with both the reference and hypothesis transcripts and asked to rate the utility of the hypothesis for answering the farmer's agricultural query on a 1--4 scale (Section~\ref{sec:llm_scoring}). Distribution of utility scores across languages:

\textbf{Hindi:}
\begin{itemize}
\item Score 4 (Excellent): 32.1\%
\item Score 3 (Acceptable): 31.3\%
\item Score 2 (Poor): 18.2\%
\item Score 1 (Unusable): 18.4\%
\end{itemize}

\textbf{Telugu:}
\begin{itemize}
\item Score 4 (Excellent): 45.2\%
\item Score 3 (Acceptable): 27.8\%
\item Score 2 (Poor): 15.1\%
\item Score 1 (Unusable): 11.9\%
\end{itemize}

\textbf{Odia:}
\begin{itemize}
\item Score 4 (Excellent): 24.1\%
\item Score 3 (Acceptable): 25.7\%
\item Score 2 (Poor): 28.3\%
\item Score 1 (Unusable): 21.9\%
\end{itemize}

Telugu shows higher utility scores than Hindi despite worse WER (33.2\% vs.\ 16.2\%). This discrepancy may reflect several factors: Telugu queries tend to be shorter, reducing error accumulation; Telugu errors may more often affect non-critical words; and the LLM scoring itself may be less calibrated across languages (see Limitations). Hindi exhibits a bimodal distribution, with 32.1\% excellent and 18.4\% unusable, suggesting high variance in recording quality or query complexity.

\section{Discussion}

\subsection{Key Findings}

\begin{enumerate}
\item \textbf{Language-Specific Challenges}: Odia consistently underperforms across most models, reflecting insufficient training data and model attention for low-resource languages. However, the inclusion of diarization-enabled models closes much of the gap, with Azure Diarize achieving 35.1\% WER compared to 70.7\% for Google STT without diarization.

\item \textbf{Domain-Specific Error Patterns}: Agricultural terms show recurring confusion patterns across all three languages, particularly in crop names and pest management terminology. The treemap analysis reveals these patterns are consistent across languages despite different scripts and phonologies.

\item \textbf{Metric Inadequacy}: Traditional WER fails to capture domain-specific risks. The AWWER analysis shows that models with acceptable WER (e.g., Google STT in Hindi at 16.2\%) may still make disproportionate errors on critical agricultural terms (AWWER rank 7th), while models like Gemini 2.5 Pro (Best Speaker) better preserve agricultural vocabulary.

\item \textbf{Diarization Impact}: Speaker diarization with best-speaker selection provides large WER reductions (up to 66\%) in languages with high multi-speaker rates. This is practically important for agricultural field recordings where multiple speakers are common.

\item \textbf{Audio Quality Effects}: Real-world agricultural recordings exhibit high rates of background talk and environmental noise. Odia's higher proportion of high-noise recordings (12.3\% vs Hindi's 2.8\%) partially explains its lower performance.

\item \textbf{Model Variation}: Large performance differences exist even among commercial systems, highlighting the need for domain evaluation.
\end{enumerate}

\subsection{Practical Implications}

For agricultural technology developers:

\begin{enumerate}
\item \textbf{Model Selection}: Google Speech-to-Text provides the best WER for Hindi and Telugu, but Gemini 2.5 Pro (Best Speaker) offers superior agricultural term accuracy (lowest AWWER). For Odia, Azure Diarize (Best Speaker) is the clear choice on both WER and AWWER.

\item \textbf{Diarization Benefit}: For deployments involving multi-speaker recordings (common in agricultural extension), enabling speaker diarization and selecting the primary speaker's transcript can reduce WER upto 66\%, depending on the multi-speaker proportion.

\item \textbf{Post-Processing Needs}: All systems require domain-specific correction mechanisms, particularly for critical agricultural terminology. The AWWER delta analysis identifies which models need more aggressive post-processing for agricultural terms.

\item \textbf{Deployment Considerations}: Hindi and Telugu can support semi-automated workflows with top models achieving $<$20\% and $<$35\% WER respectively. Odia systems, while considerably improved with diarization, still require oversight for critical agricultural advisory content.
\end{enumerate}

\subsection{Limitations}

This study has several limitations.

\textbf{Data source.} All recordings originate from a single platform (Farmer.Chat), which may not represent the full diversity of agricultural voice interactions in India. Acoustic conditions, user demographics, and query types may differ on other platforms or in other agricultural contexts.

\textbf{Ground truth quality.} Reference transcripts were produced by human annotators, but we did not measure inter-annotator agreement. Transcription of noisy, domain-specific audio in Indian languages is inherently subjective, and annotator errors would affect all reported metrics.

\textbf{LLM-based scoring.} The utility scoring relies on GPT-4o, which introduces its own biases. We did not validate LLM scores against human expert judgments, and the scoring prompt was not ablated.

\textbf{AWWER weighting.} The four-tier weighting scheme (1--4) was defined by the authors based on domain knowledge, with the lexicon derived from human-annotated reference transcripts. The specific weight values and term assignments have not been validated through ablation studies or independent expert review. Different weighting choices could alter model rankings.

\textbf{Model coverage.} Not all 10 systems were evaluated on all three languages due to language support limitations. Comparisons across languages should account for the different model subsets available.

\section{Future Work}

\begin{enumerate}
\item \textbf{Extended Language Coverage}: Evaluation of additional Indian languages (Tamil, Kannada, Malayalam).

\item \textbf{Domain Expansion}: Specialized evaluation frameworks for veterinary science, precision agriculture, and agricultural finance.

\item \textbf{Real-time Systems}: Evaluation of streaming ASR performance in agricultural contexts.

\item \textbf{Acoustic Robustness}: Assessment under varying acoustic conditions typical in agricultural settings, building on the audio quality characterization presented here.

\item \textbf{Domain-Specific Model Fine-Tuning}: Building on the agricultural terminology and error patterns identified in this study to fine-tune ASR models through domain-specific vocabulary augmentation, phonetic discrimination training, and AWWER-style priority term weighting.

\item \textbf{Open-Source Dataset and Toolkit Release}: We have released the agricultural ASR benchmark dataset (10,864 audio-transcript pairs across Hindi, Telugu, and Odia) on HuggingFace \cite{digigreen_agri_stt}. We further plan to release the domain confusion matrix and the AWWER evaluation toolkit under open licenses to support reproducible research and community-driven improvements.

\item \textbf{Community Collaboration Framework}: Establishing contribution guidelines covering audio quality standards, transcription accuracy thresholds, privacy protection for farmer voice data, and expanded regional and dialectal coverage.
\end{enumerate}

\section{Conclusion}

This study benchmarks 10 ASR models across Hindi, Telugu, and Odia for agricultural advisory use, introducing domain-specific metrics (AWWER and LLM-based utility scoring) alongside traditional WER, CER, and MER.

First, the best model varies by language and by metric: Google STT leads on WER for Hindi and Telugu, but Gemini 2.5 Pro (Best Speaker) ranks first on AWWER for Hindi, and Azure Diarize (Best Speaker) leads on both metrics for Odia. This means model selection must be guided by the deployment language and whether general accuracy or domain-term fidelity is the priority. Second, speaker diarization with best-speaker selection yields WER reductions of up to 66\%, making it the single most impactful post-processing step for multi-speaker field recordings. Third, the gap between WER and AWWER rankings exposes a blind spot in standard evaluation: models that score well on WER may disproportionately err on the agricultural terms that matter most.

These results also highlight open problems. Odia performance remains weak without diarization (70.7\% WER for several models), indicating that low-resource language support lags behind deployment needs. The AWWER weighting scheme, while useful, requires validation through expert review and ablation. LLM-based utility scoring shows promise but needs calibration against human judgment, particularly given the unexplained cross-language inconsistencies observed.

The benchmarks, error analyses, and evaluation tools presented here can serve as a starting point for agricultural ASR development in Indian languages.

\section*{Disclosure}
The authors are employees of Digital Green, which develops and operates the Farmer.Chat platform from which the evaluation data were collected. The ASR models evaluated are third-party systems with no financial relationship to Digital Green.

\end{document}